\documentclass[12pt]{article}
\usepackage{amssymb}
\usepackage{amsmath}
\usepackage{amsfonts}
\begin{document}

\author{Vladimir K. Petrov\thanks{ E-mail address: vkpetrov@yandex.ru}}
\title{{\LARGE Some peculiarities of transition from discrete to continuum Fourier
series in lattice theories}}
\date{\textit{N. N. Bogolyubov Institute for Theoretical Physics}\\
\textit{\ National Academy of Sciences of Ukraine}\\
\textit{\ 252143 Kiev, Ukraine. 11.12.2002}}
\maketitle

\begin{abstract}
Transition from discrete to continuous Fourier series is studied for the
functions becoming singular in the transition. Conditions are specified when
summing replacement by integration is inadmissible.

\end{abstract}

\section{Introduction}

Fourier series in discrete variables, or discrete Fourier series, is a routine
instrument of computation on finite lattices with cyclic border conditions.
Quite often it becomes necessary to remove lattice regularization with lattice
length $N$ approaching infinity. Ordinarily lattice theory operates with
regular functions. However, these functions may include terms which, in fact,
play the role of a regularization parameter. In other words, even when such
functions are analytical in the whole physical region for any finite $N$, the
transition to $N\rightarrow\infty$ may convert regular functions into
distributions. For the function with such nonstandard behavior, ordinary
replacement of summing by integration for $N\rightarrow\infty$ may appear inadmissible.

In this paper we focus our attention on a frequently occurring case, when such
``regularization'' terms are inversely proportional to the lattice length. The
conditions are specified when such replacement may or may not be done and
instances are discussed, when this replacement is definitely erroneous and
relevant corrections are needed. We also suggest two different methods for
correction computation.

\section{Some Remarks on Discrete Fourier Series}

In the previous work \cite{me} we considered the consequences of cyclization
procedure
\begin{equation}
\widetilde{\digamma}\left(  \varphi\right)  =\sum_{n=-\infty}^{\infty}%
\digamma\left(  \varphi+2\pi n\right) \label{per}%
\end{equation}
application to the function $\digamma\left(  \varphi\right)  $ of the
continuous variable $\varphi$. Similar procedure may be applied to the
function $F_{k}$ of a discrete variable $k=0,\pm1,\pm2,...$. We assume that
$F_{k}$ is a tempered function, i.e. that there exists a constant $c<\infty$
so that
\begin{equation}
\left|  F_{k}\right|  <\left|  k\right|  ^{c}.
\end{equation}
for $\left|  k\right|  \rightarrow\infty$

To simplify the problem, we associate $F_{k}$ with some periodic distribution
\begin{equation}
\widetilde{F}\left(  \varphi\right)  =\widetilde{F}_{+}\left(  \varphi
+i\varepsilon\right)  -\widetilde{F}_{-}\left(  \varphi-i\varepsilon\right)
\label{vla-i}%
\end{equation}
where $\varepsilon$ is a routine positive infinitesimal parameter and
$\widetilde{F}_{\pm}\left(  \varphi\right)  $ are analytical functions in
upper/lower complex half-plane $\varphi$
\begin{equation}
\widetilde{F}_{+}\left(  \varphi\right)  =\sum_{k=0}^{\infty}F_{k}\exp\left\{
ik\varphi\right\}  ;\quad\widetilde{F}_{-}\left(  \varphi\right)
=-\sum_{k=-\infty}^{-1}F_{k}\exp\left\{  ik\varphi\right\} \label{v-v}%
\end{equation}
Bearing in mind that a distribution is tempered, if and only if it is a finite
order derivative of some continuous tempered function and that any tempered
distribution may be presented in the form $\left(  \ref{vla-i}\right)  $ (see
e.g. \cite{gel-shil,bremermann,vladimirov}). Moreover, every periodic
distribution is tempered \cite{donoghue}.

As it is seen from $\left(  \ref{v-v}\right)  $ the expression $\left(
\ref{vla-i}\right)  $ coincides with an Abel-Poisson regularization of Fourier
series
\begin{equation}
\widetilde{F}\left(  \varphi\right)  =reg\sum_{k=-\infty}^{\infty}%
F_{k}e^{i\varphi k}\equiv\sum_{k=-\infty}^{\infty}F_{k}e^{i\varphi
k-\varepsilon\left|  k\right|  }\label{f-d}%
\end{equation}

Since Fourier coefficients $F_{k}$ in $\left(  \ref{f-d}\right)  $ coincide
with the considered discrete function, they are related to $\widetilde
{F}\left(  \varphi\right)  $ by the standard equation
\begin{equation}
F_{k}=\frac{1}{2\pi}\int_{-\pi+\sigma}^{\pi+\sigma}\widetilde{F}\left(
\varphi\right)  \exp\left\{  -i\varphi k\right\}  d\varphi\label{p-f}%
\end{equation}
where an arbitrary real number $\sigma$ is usually chosen as $\sigma=0$ or
$\sigma=\pi$ for convenience reasons.

Now one can easily check that applying cyclization procedure to $F_{k}$ in
$\left(  \ref{p-f}\right)  $ with the period $N$%
\begin{equation}
\widetilde{F}_{k}=\sum_{v=-\infty}^{\infty}F_{k+vN}.\label{pr}%
\end{equation}
and making allowance of the Poisson relation
\begin{equation}
\sum_{k=0}^{\infty}e^{ik\phi}\equiv2\pi\sum_{r=-\infty}^{\infty}\delta\left(
\phi-2\pi r\right) \label{del-P}%
\end{equation}
one gets
\begin{equation}
\widetilde{F}_{k}=\frac{1}{N}\sum_{n=-\infty}^{\infty}\int_{-\pi+\sigma}%
^{\pi+\sigma}\widetilde{F}\left(  \varphi\right)  \delta\left(  \varphi
-\frac{2\pi n}{N}\right)  \exp\left\{  -i\varphi k\right\}  d\varphi
\label{product}%
\end{equation}

From $\left(  \ref{product}\right)  $ we see that a sufficient condition for
$\widetilde{F}_{k}$ to exist is the existence of the product of distributions
$\widetilde{F}\left(  \varphi\right)  $ and $\delta\left(  \varphi N-2\pi
n\right)  $. Unfortunately, for an arbitrary distribution (in our case for the
arbitrary $\widetilde{F}\left(  \varphi\right)  $) such product is not defined
\cite{Schwartz,bremermann,vladimirov}, which requires\ a more detailed consideration.

It is known that Dirac $\delta$-function is rigorously defined as functional
$\left\langle \delta\left(  \varphi\right)  ,\phi\left(  \varphi\right)
\right\rangle =\phi\left(  0\right)  $ on the space $C^{0}$ of continuous
functions $\phi\left(  \varphi\right)  \in C^{0}$ \cite{vladimirov}.
Therefore, if $\widetilde{F}\left(  \varphi\right)  \in C^{0}$ then
$\widetilde{F}\left(  \varphi\right)  \phi\left(  \varphi\right)  \in C^{0}$,
i.e. $\widetilde{F}\left(  \varphi\right)  $ is a multiplicator in $C^{0}$ so
for such $\widetilde{F}\left(  \varphi\right)  $ the considered product is well-defined.

It is hardly worth mentioning that since $\widetilde{F}\left(  \varphi\right)
$ is given by the Fourier series $\left(  \ref{f-d}\right)  $ this function is
not absolutely arbitrary and had to obey specific conditions. However, even
regarded as too restrictive, Dirichlet conditions require function
$\widetilde{F}\left(  \varphi\right)  $ only to be piecewise monotonic,
piecewise differentiable and bounded for $0\leq\varphi<2\pi$ (see e.g.
\cite{whit-wats}). Under given conditions Fourier series $\left(
\ref{f-d}\right)  $ converge to $\widetilde{F}\left(  \varphi\right)  $ at all
points of $\left[  0,2\pi\right]  ,$ but at some points $\varphi_{\nu}$ of
this segment it may have first order discontinuity. Despite the fact that
$\widetilde{F}\left(  \varphi\right)  $ is defined in all these points as
$\widetilde{F}\left(  \varphi_{\nu}\right)  =\frac{1}{2}\left(  \widetilde
{F}\left(  \varphi_{\nu}-0\right)  +\widetilde{F}\left(  \varphi_{\nu
}+0\right)  \right)  $ it is no remedy for the discontinuity. Therefore, when
such discontinuity points coincide with $\varphi=2\pi n/N$, special
consideration is necessary. The distribution theory allows to consider even
more singular function $\widetilde{F}\left(  \varphi\right)  $.

If $\widetilde{F}\left(  \varphi\right)  $ is singular, then function
$\widetilde{F}_{k}$ defined in $\left(  \ref{product}\right)  $ may appear to
be nonperiodic, i.e. $\widetilde{F}_{k}\neq\widetilde{F}_{k+N}$. This case is
very similar to one considered in \cite{me} where the procedure $\left(
\ref{per}\right)  $ was applied to a function, the Fourier integral of which
has singularity. Therefore we confine ourselves to a simple case and take an
example when $\widetilde{F}\left(  \varphi\right)  $ has the simple pole
\begin{equation}
\widetilde{F}\left(  \varphi\right)  =\frac{1}{1-\exp\left\{  i\varphi
\right\}  }\label{s-pol}%
\end{equation}

From $\left(  \ref{product}\right)  $ we get
\begin{equation}
\widetilde{F}_{k}=\frac{1}{N}\sum_{n=-\infty}^{\infty}\int_{-\pi}^{\pi}%
\frac{1}{1-\exp\left\{  i\varphi\right\}  }\delta\left(  \varphi-\frac{2n\pi
}{N}\right)  \exp\left\{  -i\varphi k\right\}  d\varphi
\end{equation}

Without loss of generality we may take $N$ as an odd number
\begin{align}
\widetilde{F}_{k}  & =\frac{1}{N}\int_{-\pi}^{\pi}\frac{\exp\left\{  -i\varphi
k\right\}  }{-i+\frac{1}{2}\varphi+\frac{1}{6}i\varphi^{2}+O\left(
\varphi^{3}\right)  }\frac{1}{\varphi}\delta\left(  \varphi\right)
d\varphi\label{ex-1}\\
& +\frac{1}{N}\sum_{n=1}^{\frac{N-1}{2}}\left(  \frac{\exp\left\{
-i\frac{2n\pi}{N}k\right\}  }{1-\exp\left\{  -i\frac{2n\pi}{N}\right\}
}+\frac{\exp\left\{  i\frac{2n\pi}{N}k\right\}  }{1-\exp\left\{  i\frac{2n\pi
}{N}\right\}  }\right)  .\nonumber
\end{align}
Making allowance of the fact that for any function $f\left(  \varphi\right)  $
regular and nonvanishing in the vicinity of $\varphi=0$ we may write
\begin{equation}
f\left(  \varphi\right)  \frac{1}{\varphi}\delta\left(  \varphi\right)
=-f\left(  \varphi\right)  \delta^{\prime}\left(  \varphi\right)  =f^{\prime
}\left(  0\right)  \delta\left(  \varphi\right)  ,
\end{equation}
we easily get from $\left(  \ref{ex-1}\right)  $%

\begin{equation}
\widetilde{F}_{k}=\frac{2k+1}{N}+\frac{1}{N}\sum_{n=1}^{\frac{N-1}{2}}%
\frac{\cos\frac{2\pi nk}{N}-\cos\frac{2\pi n\left(  1-k\right)  }{N}}%
{2\sin^{2}\frac{\pi n}{N}}\label{sing}%
\end{equation}

It is evident that the second term in the right part of the equality $\left(
\ref{sing}\right)  $ is invariant under substitution $k\rightarrow k+N$. Such
term appears as the sum of contributions of the points $\varphi=2\pi n/N$;
$n\neq0$ the first term being the contribution of the pole of function
$\widetilde{F}\left(  \varphi\right)  $ at $\varphi=0$ (i.e. $n=0$). This term
is evidently nonperiodic.

When function $\widetilde{F}\left(  \varphi\right)  $\ is bounded at all
$\varphi=2\pi n/N$\ and continuous in the vicinities of those points, we
regard that the sufficient condition for $\widetilde{F}_{k}$ to be periodic is
fulfilled. Indeed, from $\left(  \ref{product}\right)  $ we obtain
\begin{equation}
\widetilde{F}_{k}=\frac{1}{N}\sum_{n=0}^{N-1}\widetilde{F}\left(  \frac{2\pi
n}{N}\right)  \exp\left\{  i\frac{2\pi n}{N}k\right\} \label{dd}%
\end{equation}
Therefore, since $\widetilde{F}_{k}$ is presented as the finite Fourier series
it is an evidently periodic function with the period $N$. It is clear that
$\widetilde{F}\left(  \frac{2\pi n}{N}\right)  $, as a function of discrete
variable $n$, is periodic with the same period by definition.

Making allowance of the simple identity
\begin{equation}
\frac{1}{N}\sum_{k=0}^{N-1}\exp\left\{  -i\frac{2\pi k}{N}n\right\}
=\sum_{v=-\infty}^{\infty}\delta_{n}^{vN}\label{ort}%
\end{equation}
one may write the transform inverse to $\left(  \ref{dd}\right)  $
\begin{equation}
\widetilde{F}\left(  \frac{2\pi n}{N}\right)  =\sum_{k=0}^{N-1}\widetilde
{F}_{k}\exp\left\{  -i\frac{2\pi n}{N}k\right\} \label{dd-}%
\end{equation}

We see that $\left(  \ref{dd-}\right)  $ is nothing but the discrete version
of Fourier series $\left(  \ref{f-d}\right)  $, as well as $\left(
\ref{dd}\right)  $ is the discrete version of $\left(  \ref{p-f}\right)  $.

It is clear that the above condition on $\widetilde{F}\left(  \varphi\right)
$\ may be related to the convergence of series $\left(  \ref{f-d}\right)
$\ or rather to the behavior $F_{k}$\ for $k\rightarrow\pm\infty$ that, in its
turn, is related to the convergence of series in $\left(  \ref{pr}\right)  $.
For instance, necessary conditions on $\widetilde{F}\left(  \varphi\right)
$\ are fulfilled when series $\left(  \ref{f-d}\right)  $\ is uniformly
convergent on a closed segment $-\pi\leq\varphi\leq\pi$ (see e.g.
\cite{whit-wats}).

However, one may impose on $F_{k}$ a weaker restriction and consider
$\widetilde{F}\left(  \varphi\right)  $ which has a limited variation, i.e.
such functions which for arbitrary partition $-\pi\leq\varphi_{v}%
<\varphi_{v+1}\leq$\ $\pi$\ and arbitrary $N$\ obey
\begin{equation}
\sum_{v=0}^{N-1}\left|  \widetilde{F}\left(  \varphi_{v}\right)
-\widetilde{F}\left(  \varphi_{v+1}\right)  \right|  <M.
\end{equation}
for some $M<\infty$.

A function with a limited variation in a segment is obviously bounded on it.
Moreover, as it is shown by\ N.Weiner (see e.g. \cite{edwards}),
such\ function is continuous, if (and only if)
\begin{equation}
\lim_{K\rightarrow\infty}\frac{1}{K}\sum_{k=-K}^{K}\left|  kF_{k}\right|
=0.\label{W}%
\end{equation}

This condition, however, may be regarded as too restrictive as well, just for
that very reason that it claims 'good' behavior of $\widetilde{F}\left(
\varphi\right)  $ for all $\varphi$, whereas it is actually needed only at
infinitesimal vicinities of $\varphi=2\pi n/N$. Hence there is no reason to
expect that each particular $\widetilde{F}\left(  \varphi\right)  $ that is
presented as discrete Fourier series $\left(  \ref{dd-}\right)  $ obligatory
obeys such conditions .

\section{Regularization Removal and Transition to Continuum}

The discrete Fourier series $\left(  \ref{dd}\right)  $ and $\left(
\ref{dd-}\right)  $ are commonly used in lattice theories with periodic
boundary conditions. Quite often one needs to apply a procedure inverse to
$\left(  \ref{pr}\right)  $, which involves the removal of the lattice
regularization and the lattice length $N$ then tends to infinity.

This section considers peculiarities of the transition from discrete to
continuous Fourier series with $N\rightarrow\infty$. As it is said above, we
are interested in a very particular aspect of such a fundamental problem as
transition to continuum in a lattice gauge theory, and namely in the
transition to continuum of the discrete Fourier series for functions which are
regular at finite $N$, but convert into distributions with $N\rightarrow
\infty$. Naturally, some important issues of transition to continuum are
outside the limits of this study, nonetheless, the suggested approach allows,
in particular, to compute the fermion contribution on an extremely anisotropic
lattice \cite{det}.

Let $\widetilde{F}\left(  \varphi\right)  $ in $\left(  \ref{dd}\right)  $
obey the conditions of Lebesgue theorem, i.e. $\widetilde{F}\left(
\varphi\right)  $ is a bounded function with a set of measure zero
discontinuity points. In this case $\widetilde{F}\left(  \varphi\right)  $ is
Riemann integrable, i.e. the difference between upper and lower sum Darboux
vanishes for arbitrary partition $\left(  -\pi\leq\varphi_{v}<\varphi
_{v+1}\leq\mathrm{\ }\pi;\quad\nu=0,1,...,N-1\right)  $ when the number of
terms $N$ infinitely increases\footnote{For piecewise-monotonic $\widetilde
{F}\left(  \varphi\right)  $ the sum in $\left(  \ref{dd}\right)  $
decomposees into partial sums over monotonic pieces. Each partial sum
coinsides either with upper or with lower sum Darboux.}. For $N\rightarrow
\infty$ this allows us to make a change in $\left(  \ref{dd}\right)  $
\begin{equation}
\frac{2\pi n}{N}\rightarrow\varphi;\qquad\frac{1}{N}\sum_{n=0}^{N-1}%
\rightarrow\frac{1}{2\pi}\int_{0}^{2\pi}d\varphi;\label{lim}%
\end{equation}
which transforms $\left(  \ref{dd}\right)  $ into $\left(  \ref{p-f}\right)  $.

For any finite value $s$ we may put $s/N\rightarrow0$. If, however, $s=0$ is a
singular point of function $\widetilde{F}\left(  \varphi\right)  $ we need to
preserve the information about the direction in which $s/N$ approach zero in a
complex $s$-plane%

\begin{equation}
s/N\rightarrow\varepsilon\exp\left\{  i\arg s\right\}
\end{equation}

When, however, a transition for $\widetilde{F}_{k}$ in $\left(  \ref{dd-}%
\right)  $ is of interest, one needs $\widetilde{F}_{k}$ to obey the
conditions of Lebesgue theorem. If it is the case, then transition to
continuous variable $2\pi k/N\rightarrow\varphi$ may be fulfilled by the
following substitution
\begin{equation}
\frac{1}{N}\sum_{k=0}^{N-1}\rightarrow\frac{1}{2\pi}\int_{0}^{2\pi}%
d\varphi;\quad\widetilde{F}_{k}\rightarrow\widetilde{F}\left(  \varphi\right)
;\quad\widetilde{F}\left(  \frac{2\pi}{N}n\right)  \rightarrow F_{n}%
.\label{lim-2}%
\end{equation}

To estimate the error, introduced by the procedure$\left(  \ref{lim}\right)  $
we apply the Poisson formula to truncated functions%

\begin{equation}
\overline{F}\left(  \varphi\right)  =\left\{
\begin{array}
[c]{ccc}%
\widetilde{F}\left(  \varphi\right)  ; &  & 0\leq\varphi<2\pi;\\
0; &  & otherwise.
\end{array}
\right. \label{trunc}%
\end{equation}
that gives
\begin{equation}
\frac{1}{N}\sum_{n=0}^{N-1}\widetilde{F}\left(  \frac{2\pi n}{N}\right)
=\sum_{k=-\infty}^{\infty}\frac{1}{2\pi}\int_{0}^{2\pi}\widetilde{F}\left(
\varphi\right)  \exp\left\{  -i\varphi kN\right\}  d\varphi\label{poiss}%
\end{equation}
so, taking into account the definition $\left(  \ref{p-f}\right)  $ we may
write
\begin{equation}
\frac{1}{N}\sum_{n=0}^{N-1}\widetilde{F}\left(  \frac{2\pi n}{N}\right)
=\sum_{k=-\infty}^{\infty}F_{kN}=F_{0}+\Delta\label{int}%
\end{equation}

It is clear that the application of procedure $\left(  \ref{lim}\right)  $ to
the left part in $\left(  \ref{int}\right)  $ gives but the term $F_{0}$.
Therefore, to obtain an exact result, one had to include the correction
$\Delta$ which is defined as
\begin{equation}
\Delta=\sum_{k\neq0}F_{kN}=\sum_{k=1}^{\infty}\left(  F_{kN}+F_{-kN}\right)
\label{DEL}%
\end{equation}

For $\Delta$\ to vanish, it is sufficient that for all $k$%

\begin{equation}
\widetilde{F}\left(  \varphi\right)  \exp\left\{  -i\varphi kN\right\}
\rightarrow0;\quad N\rightarrow\infty\label{fa -}%
\end{equation}
Since $kN$ is an integer number we may rewrite $\left(  \ref{fa -}\right)  $
as
\begin{equation}
\widetilde{F}\left(  \varphi\right)  \exp\left\{  -i\left(  \varphi-\sigma
_{0}\right)  kN\right\}  \rightarrow0;\quad N\rightarrow\infty\label{fa}%
\end{equation}
where $\sigma_{0}=2\pi n_{0}/N$ and $n_{0}$ is an arbitrary integer number. In
other words, the result of application $\left(  \ref{lim}\right)  $ to
left-hand member of $\left(  \ref{poiss}\right)  $ coincides with right-hand
one only if the condition $\left(  \ref{fa}\right)  $ is satisfied for any
$\sigma_{0}$ and all $k\neq0$.

According to Riemann-Lebesgue lemma (see. e.g. \cite{whit-wats}) all $F_{kN}%
$\ in $\left(  \ref{int}\right)  $\ with $k$\ $\neq0$\ turn into zero with
$N\rightarrow\infty$, if either the integral in $\left(  \ref{poiss}\right)  $
\ converge absolutely for $k=0$\ or $\widetilde{F}\left(  \varphi\right)
$\ has limited variation on $0\leq\varphi<2\pi$.

For singular enough $\widetilde{F}\left(  \varphi\right)  $, the conditions of
Lebesgue theorem may be violated and\ procedure $\left(  \ref{lim}\right)
$\ transforms the sum into improper integral, which value will depend on the
choice of the way singularity is approached. The result of proceeding to limit
in expressions like $\left(  \ref{fa -}\right)  $ was studied in detail
\cite{brych,br-shir}. Various type singularities of $\widetilde{F}\left(
\varphi\right)  $, located infinitely near to $\varphi=0$ were
considered.$\allowbreak$ Results obtained in \cite{brych,br-shir} allow to
conclude that the condition $\left(  \ref{fa}\right)  $ is satisfied, if
$\widetilde{F}\left(  \varphi\right)  $ is regular at infinitesimal vicinities
of all $\varphi=2\pi n/N$, where $n=0,...,N-1$.

Commonly the function $\widetilde{F}\left(  \varphi\right)  $\ in $\left(
\ref{fa}\right)  $\ doesn't depend on $N$. In a lattice gauge theory, however,
$\widetilde{F}\left(  \varphi\right)  $\ may include small parameters, such as
fermion mass term $ma$, which tends to zero with $N\rightarrow\infty$,
provided that inverse temperature $1/T=aN$\ remains finite. Actually, it
introduces\ a dependence on $N$ into $\widetilde{F}\left(  \varphi\right)  $.
Therefore, it is not without interest to consider a case when $\widetilde
{F}\left(  \varphi\right)  $ includes some 'regularization' term which is
proportional to $1/N$, i.e. singularity of $\widetilde{F}\left(
\varphi\right)  $ is located in a complex $\varphi$ plane at finite distance
$\varphi_{s}\sim$ $i/N$ from the real axis. In this case the increase of
exponent oscillations in $\left(  \ref{fa}\right)  $ is matches singular point
$\varphi_{s}$ approaching physical region. In Appendix we suggest a simple
recipe to modify the results obtained in \cite{brych,brych-prud} for this case.

\section{Examples of nonstandard behavior}

Let us consider an example of a periodic function with the simple pole
\begin{equation}
\widetilde{F}\left(  \frac{2\pi n}{N}\right)  =\left(  1-\exp\left\{
i\frac{2\pi n}{N}-\frac{\varkappa}{N}\right\}  \right)  ^{-1}\exp\left\{
-m\left(  i\frac{2\pi n}{N}-\frac{\varkappa}{N}\right)  \right\} \label{f-pol}%
\end{equation}
It is clear that the role of regularization term in $\left(  \ref{f-pol}%
\right)  $ is played by $\varkappa/N$, where $\varkappa$ is finite and
$\operatorname{Im}\varkappa=0$.

Writing series expansion
\begin{equation}
\left(  1-e^{\frac{2\pi in-\varkappa}{N}}\right)  ^{-1}=\theta\left(
-\varkappa\right)  +\sum_{r=0}^{\infty}\left(  \theta\left(  \varkappa\right)
e^{r\frac{2\pi in-\varkappa}{N}}-\theta\left(  -\varkappa\right)
e^{-r\frac{2\pi in-\varkappa}{N}}\right)
\end{equation}
and collecting coefficients at $r=k$ modulo $N$, we obtain
\begin{equation}
\left(  1-e^{\frac{2\pi in-\varkappa}{N}}\right)  ^{-1}=\theta\left(
-\varkappa\right)  +\frac{\theta\left(  \varkappa\right)  }{1-e^{-\varkappa}%
}\sum_{k=0}^{N-1}e^{k\frac{2\pi in-\varkappa}{N}}-\frac{\theta\left(
-\varkappa\right)  }{1-e^{\varkappa}}\sum_{k=0}^{N-1}e^{-k\frac{2\pi
in-\varkappa}{N}};\label{ex}%
\end{equation}

As it follows from$\left(  \ref{ort}\right)  $ one may write
\begin{equation}
\frac{1}{N}\sum_{n=0}^{N-1}\exp\left\{  -\frac{2\pi in}{N}\left(  m+k\right)
\right\}  =\delta_{N}^{m+k}\left(  1-\delta_{0}^{m}\right)  +\delta_{0}%
^{m}\delta_{0}^{k}%
\end{equation}
for nonnegative integer $m<N$ and $n<N$. Therefore, from $\left(
\ref{ex}\right)  $ we finally get
\begin{equation}
\frac{1}{N}\sum_{n=0}^{N-1}\left(  1-e^{\frac{2\pi in-\varkappa}{N}}\right)
^{-1}e^{-m\frac{2\pi in-\varkappa}{N}}=\frac{1}{1-e^{-\varkappa}}%
-\theta\left(  -\varkappa\right)  \left(  1-\delta_{0}^{m}\right) \label{pol}%
\end{equation}
Proceeding to limit $N\rightarrow\infty$ in $\left(  \ref{pol}\right)  $ is
trivial, because the right-hand member doesn't depend on $N$.

On the other hand, the application of procedure $\left(  \ref{lim}\right)  $
to the left-hand member in $\left(  \ref{pol}\right)  $ gives%

\begin{equation}
\frac{1}{2\pi}\int_{-\pi}^{\pi}\frac{\exp\left\{  -i\varphi m\right\}
}{1-\exp\left\{  i\varphi-\varepsilon\operatorname*{signum}\left(
\varkappa\right)  \right\}  }d\varphi=\operatorname*{signum}\left(
\varkappa\right)  +\theta\left(  -\varkappa\right)  \delta_{m}^{0}%
\label{ex-lim}%
\end{equation}

We see that the procedure $\left(  \ref{lim}\right)  $ leads to a result which
differs essentially from an exact one given in $\left(  \ref{pol}\right)  $.
The reason is that the difference between upper and lower sum Darboux computed
for $\widetilde{F}\left(  2\pi in/N\right)  $ in $\left(  \ref{pol}\right)  $
doesn't vanish for $N\rightarrow\infty$, thus the sum in $\left(
\ref{pol}\right)  $ doesn't convert into Riemannian integral when
$N\rightarrow\infty$.\ It may be anticipated, because Riemannian integral is
defined only on bounded function, whereas $\left(  \ref{pol}\right)  $ remain
finite only for finite $N$.

Let us compute now the correction $\Delta$ defined in $\left(  \ref{DEL}%
\right)  .$ In considered case $F_{kN}$ is given by
\begin{align}
F_{kN}  & =\int_{0}^{2\pi}\frac{\exp\left\{  -i\varphi kN\right\}  }%
{1-\exp\left\{  i\varphi-\frac{\varkappa}{N}\right\}  }\frac{d\varphi}{2\pi
}\nonumber\\
& =\left[  \theta\left(  \varkappa\right)  \theta\left(  k\right)
+\theta\left(  -\varkappa\right)  \delta_{kN}^{-m}-\theta\left(
-\varkappa\right)  \theta\left(  -k\right)  \right]  \exp\left\{
-k\varkappa\right\}
\end{align}
and we easily get
\begin{equation}
\Delta=\frac{1}{1-e^{-\varkappa}}-\theta\left(  \varkappa\right) \label{cor}%
\end{equation}
Thereby it is confirmed that the correction $\left(  \ref{cor}\right)  $ added
to $\left(  \ref{ex-lim}\right)  $ allows to obtain the exact result $\left(
\ref{pol}\right)  $.

Applying the differentiation with respect to a parameter $\varkappa$ one may
easily repeat the computation for arbitrary rational function of $\exp\left\{
-2\pi in/N\right\}  $. Previously considered case may be easily generalized in
another way, if we consider
\begin{equation}
F\left(  \varphi\right)  =\int_{-\infty}^{\infty}f\left(  \varkappa\right)
\frac{1}{1-\exp\left\{  i\varphi-\frac{\varkappa}{N}\right\}  }d\varkappa
\end{equation}
where the integral has to be interpreted in the sense of a Cauchy principal
value. Making allowance for $\left(  \ref{pol}\right)  $\ we get
\begin{equation}
\frac{1}{N}\sum_{n=0}^{N-1}F\left(  \frac{2\pi n}{N}\right)  \exp\left\{
-\frac{2\pi in}{N}m\right\}  =\int_{-\infty}^{\infty}\frac{f\left(
\varkappa\right)  }{1-e^{-\varkappa}}d\varkappa-\left(  1-\delta_{m}%
^{0}\right)  \int_{-\infty}^{0}f\left(  \varkappa\right)  d\varkappa
\label{gen}%
\end{equation}

On the other hand, the procedure $\left(  \ref{lim}\right)  $\ applied to the
left-hand member in $\left(  \ref{gen}\right)  $\ leads to%

\begin{align}
& \int_{-\pi}^{\pi}\left(  \int_{-\infty}^{\infty}\frac{f\left(
\varkappa\right)  d\varkappa}{1-\exp\left\{  i\varphi-\varepsilon
\operatorname*{signum}\left(  \varkappa\right)  \right\}  }\right)
e^{-i\varphi m}\frac{d\varphi}{2\pi}\nonumber\\
& =\int_{0}^{\infty}f\left(  \varkappa\right)  d\varkappa-\left(  1-\delta
_{m}^{0}\right)  \int_{-\infty}^{0}f\left(  \varkappa\right)  d\varkappa
\end{align}
which disagrees, as it may be anticipated, with $\left(  \ref{gen}\right)  $.

Let us now consider as an example the periodic function with logarithmic
singularity. From ( \cite{pbm}6.1.1$\left(  10\right)  $) one can easily get%

\begin{equation}
\prod_{n=0}^{N-1}\left(  1-\exp\left\{  \frac{2\pi in-\varkappa}{N}\right\}
\right)  =1-e^{-\varkappa}%
\end{equation}
that leads to
\begin{equation}
\sum_{n=0}^{N-1}\ln\left(  1-\exp\left\{  \frac{2\pi in-\varkappa}{N}\right\}
\right)  =\ln\left(  1-e^{-\varkappa}\right) \label{ex-2}%
\end{equation}
On the other hand, the procedure $\left(  \ref{lim}\right)  $ gives
\begin{equation}
\frac{1}{2\pi}\int_{-\pi}^{\pi}\ln\left(  1-\exp\left\{  i\left(
\varphi+i\varepsilon\operatorname*{signum}\left(  \varkappa\right)  \right)
\right\}  \right)  d\varphi=0
\end{equation}
that evidently disagree with $\left(  \ref{ex-2}\right)  $ at finite
$\varkappa$.

In some instances the procedure $\left(  \ref{lim}\right)  $ leads to
ambiguous results and becomes inadmissible. Indeed, from (\cite{pbm}%
4.4.5$\left(  7\right)  $) one can get
\begin{equation}
Q_{N}\left(  \varkappa\right)  =\frac{1}{N}\sum_{k=0}^{N}\frac{\cos\left(
\pi\frac{k+\frac{1}{2}}{N}m\right)  \sin\left(  \pi\frac{k+\frac{1}{2}}%
{N}\right)  }{\cos\frac{\varkappa}{N}-\cos\left(  \pi\frac{k+\frac{1}{2}}%
{N}\right)  }e^{i\pi k}=\frac{\cos\left(  m\varkappa/N\right)  }{\cos
\varkappa}.
\end{equation}
Hence, one obtains a well defined result for $N\rightarrow\infty$ and any
finite $m$
\begin{equation}
\lim_{N\rightarrow\infty}Q_{N}\left(  \varkappa\right)  \equiv Q\left(
\varkappa\right)  =\frac{1}{\cos\varkappa}.
\end{equation}

On the contrary, the procedure $\left(  \ref{lim}\right)  $ gives
\begin{equation}
Q\left(  \varkappa\right)  =\lim_{N\rightarrow\infty}\frac{1}{2\pi}\int_{-\pi
}^{\pi}\frac{e^{i\varphi N}\cos\left(  \frac{\varphi}{2}m\right)  \sin\left(
\frac{\varphi}{2}\right)  }{\cos\left(  \eta\right)  -\cos\left(
\frac{\varphi}{2}\right)  }d\varphi\label{l-1}%
\end{equation}
where $\eta=\varepsilon\exp\left\{  i\arg\varkappa\right\}  $.

Computation in $\left(  \ref{l-1}\right)  $ of $Q\left(  \varkappa\right)  $
leads to an ambiguous result, because, on the one hand, $\exp\left\{  i\varphi
N\right\}  $ turns integrand into zero for all $\varphi$ except the vicinity
of $\varphi=0$, on the other, - the integrand has the pole at $\varphi=2\eta$
that is located infinitely near to $\varphi=0$. Therefore, $Q\left(
\varkappa\right)  $ includes the factor $\lim_{N\rightarrow\infty}\exp\left\{
2i\eta N\right\}  $ which cannot be defined unambiguously without additional assumptions.

\section{Appendix. Asymptotic Series for Distributions}

Here we briefly sketch some of the results obtained in \cite{brych} for the
asymptotic expansions of $\exp\left\{  \mp i\varphi N\right\}  F\left(
\varphi\right)  $ for $N\rightarrow\infty$ with necessary modifications which
we suggest for a case when the regularization term in $F\left(  \varphi
\right)  $ is proportional to $1/N$.

As it is known (see e.g. \cite{gel-shil}), temperate distribution is defined
as functional on infinitely differentiable fast decreasing test functions
$\Phi\left(  \varphi\right)  $ i.e. $\lim_{\left|  \varphi\right|
\rightarrow\infty}\left|  \varphi\right|  ^{m}\Phi^{\left(  p\right)  }\left(
\varphi\right)  =0;$ for any $m\geq0;$ $p\geq0$. As it pointed out in
\cite{brych}, one may write the asymptotic (Poincare) series for the
functional defined on any test function $\Phi\left(  \varphi\right)  $
\begin{equation}
T_{N}^{F}\equiv\int\exp\left\{  \mp i\varphi N\right\}  F\left(
\varphi\right)  \Phi\left(  \varphi\right)  d\varphi\simeq\sum_{n=0}^{\infty
}\psi_{n}^{\pm}\left(  N\right)  \int C_{n}^{\pm}\left(  \varphi\right)
\Phi\left(  \varphi\right)  d\varphi\label{poi}%
\end{equation}
where $\psi_{n}^{\pm}\left(  x\right)  $ for $x\rightarrow\pm\infty$
constitute asymptotic sequences obeying
\begin{equation}
\psi_{n+1}^{\pm}\left(  x\right)  /\psi_{n}^{\pm}\left(  x\right)  =O\left(
1/x\right)  ;\quad\frac{\partial\psi_{n}^{\pm}\left(  x\right)  }{\partial
x}/\psi_{n}^{\pm}\left(  x\right)  =O\left(  1/x\right) \label{con}%
\end{equation}
So, differentiating $n$-times $\left(  \ref{poi}\right)  $, we may get in
particular
\begin{equation}
\varphi^{n}C_{n}^{\pm}\left(  \varphi\right)  =0
\end{equation}
that means
\begin{equation}
C_{n}^{\pm}\left(  \varphi\right)  =\sum_{k=0}^{n-1}C_{nk}^{\pm}%
\delta^{\left(  k\right)  }\left(  \varphi\right) \label{C}%
\end{equation}
The simplest form of $C_{n}^{\pm}\left(  \varphi\right)  \psi_{n}^{\pm}\left(
N\right)  $ satisfying both$\left(  \ref{C}\right)  $ and condition $\left(
\ref{con}\right)  $ is%

\begin{equation}
C_{n}^{\pm}\left(  \varphi\right)  \psi_{n}^{\pm}\left(  N\right)  =\sum
_{k=0}^{n-1}C_{nk}^{\pm}\psi_{n}^{\pm}\left(  N\right)  \delta^{\left(
k\right)  }\left(  \varphi\right)  \sim\sum_{k=0}^{n-1}P_{nk}^{\pm}\left(  \ln
N\right)  N^{-r-k}\delta^{\left(  k\right)  }\left(  \varphi\right)
\label{sim}%
\end{equation}
where polynomials $P_{nk}^{\pm}\left(  \ln N\right)  $ and constant $r$ are
defined by the specific form of $F\left(  \varphi\right)  $.

Bearing in mind that any tempered (not obligatory periodic) distribution may
be presented in the form analogous to $\left(  \ref{vla-i}\right)  $ (see e.g.
\cite{vladimirov}) where $F_{\pm}\left(  \varphi\right)  $ are analytical
functions in upper/lower complex half-plane $\varphi$
\begin{equation}
F\left(  \varphi\right)  =F_{+}\left(  \varphi+i\varepsilon\right)
-F_{-}\left(  \varphi-i\varepsilon\right)
\end{equation}
it can be shown that \cite{brych}%

\begin{equation}
F_{\pm}\left(  \varphi\pm i\varepsilon\right)  \exp\left\{  \pm i\varphi
N\right\}  \sim0,\qquad N\rightarrow\infty\label{zer}%
\end{equation}
and%

\begin{equation}
F_{\pm}\left(  \varphi\pm i\varepsilon\right)  \exp\left\{  \mp i\varphi
N\right\}  \simeq\sum_{n=0}^{\infty}C_{n}^{\pm}\left(  N\right)
\delta^{\left(  n\right)  }\left(  \varphi\right)  ,\qquad N\rightarrow
\infty\label{fin}%
\end{equation}

As already mentioned, we are interested in a specific case, when singularity
of $F_{\pm}$ is located not at infinitesimal distance $\pm i\varepsilon$ but
at finite\footnote{In this section we assume $\varkappa>0$.} one $\pm
i\varkappa/N$ from the real axes of $\varphi$ and approaches the real axes
only with $N\rightarrow\infty$. Thereby, instead of a definition given in
$\left(  \ref{poi}\right)  $ we may write
\begin{equation}
T_{N}^{F}\equiv\int\exp\left\{  \mp i\varphi N\right\}  F_{\pm}\left(
\varphi\pm i\frac{\varkappa}{N}\right)  \Phi\left(  \varphi\right)
d\varphi\label{poi - 2}%
\end{equation}
and after a simple shift $\varphi\rightarrow\varphi\mp i\left(  \varkappa
/N-\varepsilon\right)  $ we obtain%

\begin{equation}
T_{N}^{F}=\int\exp\left\{  \mp i\left(  \varphi\mp i\left(  \frac{\varkappa
}{N}-\varepsilon\right)  \right)  N\right\}  F_{\pm}\left(  \varphi\pm
i\varepsilon\right)  \Phi\left(  \varphi\mp i\left(  \frac{\varkappa}%
{N}-\varepsilon\right)  \right)  d\varphi\label{shif}%
\end{equation}

The test functions $\Phi\left(  \varphi\right)  $ and their derivatives
$\Phi^{\left(  n\right)  }\left(  \varphi\right)  $ should not be obligatory
analytical, but they\ are infinitely differentiable. Hence, their Taylor
series may be taken as asymptotic one, no matter it is convergent or divergent
in the area in question \cite{fedorjuk}
\begin{equation}
\Phi^{\left(  n\right)  }\left(  \varphi\mp i\left(  \frac{\varkappa}%
{N}-\varepsilon\right)  \right)  =\sum_{m=0}^{M-1}\frac{1}{m!}\Phi^{\left(
n+m\right)  }\left(  \varphi\right)  \left(  \mp i\left(  \frac{\varkappa}%
{N}-\varepsilon\right)  \right)  ^{m}+O\left(  \left(  \frac{\varkappa}%
{N}-\varepsilon\right)  ^{M}\right) \label{asy}%
\end{equation}

Taking into account that parameter $\varepsilon$ is infinitely small in
comparison with $\varkappa/N$ we obtain from $\left(  \ref{shif}\right)  $%

\begin{equation}
T_{N}^{F}=e^{-\varkappa}\sum_{n=0}^{\infty}\sum_{m=0}^{M-1}C_{n}^{\pm}\left(
N\right)  \left(  -1\right)  ^{n}\frac{\Phi^{\left(  n+m\right)  }\left(
0\right)  }{m!}\left(  \mp\frac{i\varkappa}{N}\right)  ^{m}+O\left(
\frac{\varkappa^{M}}{N^{M}}\right)
\end{equation}
and for $N\rightarrow\infty$ we get
\begin{equation}
T_{N}^{F}=e^{-\varkappa}\sum_{n=0}^{\infty}C_{n}^{\pm}\left(  N\right)
\left(  -1\right)  ^{n}\Phi^{\left(  n\right)  }\left(  0\right)  \left(
1+O\left(  \frac{\varkappa}{N}\right)  \right) \label{br}%
\end{equation}
or
\begin{equation}
\exp\left\{  \mp i\varphi N\right\}  F_{\pm}\left(  \varphi\pm i\frac
{\varkappa}{N}\right)  \simeq\exp\left\{  -\varkappa\right\}  \sum
_{n=0}^{\infty}C_{n}^{\pm}\left(  N\right)  \delta^{\left(  n\right)  }\left(
\varphi\right)
\end{equation}
that with $\left(  \ref{fin}\right)  $ and $\left(  \ref{zer}\right)  $ gives
after change $N\rightarrow kN;\quad\varkappa\rightarrow k\varkappa$
\begin{align}
& \exp\left\{  \mp i\varphi kN\right\}  F_{\pm}\left(  \varphi\pm i\frac{k}%
{N}\right) \nonumber\\
& \simeq\left\{
\begin{array}
[c]{ccc}%
\exp\left\{  -k\varkappa\right\}  \sum_{n=0}^{\infty}C_{n}^{\pm}\left(
kN\right)  \delta^{\left(  n\right)  }\left(  \varphi\right)  & for & k>0\\
0 & for & k<0
\end{array}
\right.
\end{align}
We see that correction $\Delta$ in $\left(  \ref{int}\right)  $ may be written
in a form
\begin{align}
\Delta & =\sum_{k=1}^{\infty}\exp\left\{  -\varkappa k\right\}  \frac{1}{2\pi
}\int_{0}^{2\pi}\sum_{n=0}^{\infty}C_{n}^{\pm}\left(  kN\right)
\delta^{\left(  n\right)  }\left(  \varphi-\sigma_{0}\right)  d\varphi
\nonumber\\
& =\frac{1}{2\pi}\sum_{k=1}^{\infty}\exp\left\{  -\varkappa k\right\}
C_{0}^{\pm}\left(  kN\right) \label{DEL-B}%
\end{align}

As it is seen from $\left(  \ref{DEL-B}\right)  $ the correction $\Delta$ is
presented as a series which is a discrete version of Laplace transform, called
$Z$ - transformation (see e.g. \cite{doe}). Since we consider $\varkappa
/N\rightarrow+0$, coefficients $C_{0}^{\pm}\left(  x\right)  $ in this series
are the same as those computed for the asymptotic expansions in
\cite{brych-prud}. It should be noted, that distributions studied in
\cite{brych-prud} are nonperiodic. Hence, if direct application of the method
developed in \cite{brych-prud} to the periodic distribution under
consideration\ is technically difficult, it may be convenient to apply the
cyclization procedure considered in \cite{me} to those distributions that have
already been studied in \cite{brych-prud}.

\section{Conclusions}

Some particularities of the transition from discrete to continuous Fourier
series in a limit $N\rightarrow\infty$ are discussed. A case is studied when
the regularization parameter is proportional to $1/N$ and the considered
functions are analytical for $N<\infty$, but they are converted into
generalized functions or distributions when regularization is removed, i.e.
$N\rightarrow\infty$.

Some simple and rather general examples of functions with 'nonstandard'
behavior were presented to clarify the conditions under which ordinary
replacement of summing by integration is inadmissible and the corresponding
corrections $\Delta$ were computed. It is also pointed out that when the
periodic function\ $\widetilde{F}\left(  \varphi\right)  $ obeys the
conditions of Riemann-Lebesgue lemma, the corresponding correction$\ $turns
into zero.

We suggested a modification of asymptotic series expansions proposed in
\cite{brych,br-shir} for a case with regularization parameter proportional to
$1/N $ and applied the modified series for the computation of the correction
$\Delta$.

\end{document}